\documentclass[a4paper]{article}

\usepackage{graphicx}
\usepackage{amsmath}	
\usepackage{amssymb}	
\usepackage[round]{natbib}
\usepackage{enumitem}
\usepackage{hyperref}

\newlist{legal}{enumerate}{10}
\setlist[legal]{label*=\arabic*.}

\begin{document}
\title{Fast and Precise Light Curve Model for Transiting Exoplanets with Rings}

\author{
Edan Rein\thanks{Gymnasia Re'alit High School, Rishon LeZion, Israel}$^{\mbox{ ,}}$\thanks{The Davidson Institute of science education, Rehovot, Israel.}$^{\mbox{ ,}}$\thanks{E-mail: edanrein2000@gmail.com},
Aviv Ofir\thanks{Department of Earth and Planetary Sciences, Weizmann Institute of Science, Rehovot, 76100, Israel.}}

\maketitle

\begin{abstract}
		The presence of silicate material in known rings in the Solar System raises the possibility of ring systems existing even within the snow line -- where most transiting exoplanets are found. Previous studies have shown that the detection of exoplanetary rings in transit light curves is possible, albeit challenging. To aid such future detection of exoplanetary rings, we present the Polygon+Segments model for modelling the light curve of an exoplanet with rings. This high-precision model includes full ring geometry as well as possible ring transparency and the host star's limb darkening. It is also computationally efficient, requiring just a 1D integration over a small range, making it faster than existing techniques. The algorithm at its core is further generalized to compute the light curve of any set of convex primitive shapes in transit ({\it e.g.} multiple planets, oblate planets, moons, rings, combination thereof, etc.) while accounting for their overlaps. The python source code is made available.
\end{abstract}

keywords:
methods: data analysis -- methods: numerical -- techniques: photometric -- occultations -- planets and satellites: rings

\section{Introduction}
	
	All giant planets in the Solar System, and even some minor bodies have rings \citep{BragaRibas04, Ortiz15}. It is therefore natural to expect that exoplanetary rings exist. The various ring systems in the Solar System have varying fractions of ices (especially water ice) and rocky material \cite[][and references therein]{Esposito10,dePater17}. It was therefore speculated that even planets that are closer to their host star, even within the snow line, may host rings. \citet{Schlichting11} found that most of the observed exoplanets have sufficiently large Roche radii to support rings despite their relative proximity to their host, and speculated that such rings may allow observers to gain insight into the planet's quadruple gravitational moment and composition. 
	
    In the far future, resolved images of exoplanets may reveal the presence of extended ring systems around them. However, in the foreseeable future, detecting planetary rings will be possible only through the transit method. Unfortunately, the light curve of a planet with rings closely resembles that of a ringless planet with a similar projected area, and the detection of rings in transit hinges on the detailed analysis of the ingress/egress segments of the transit \citep{BarnesFortney04}. Critically, these very segments are also where degeneracies in the parameters of the standard ringless model \citep{MA2002} are most important, as well as stellar limb-darkening uncertainties. \citet{Neilson17} caution that the commonly used limb darkening laws induce systematic errors with an amplitude of hundreds of ppms.
    
    The possibility of detecting exoplanetary rings using the transit method was considered even before the first transiting planet was observed \citep{Schneider99}. Still, there has not been a definitive detection of a ring system to date. The two best candidates currently known are 1SWASP J140747.93-394542.6 (hereafter J1407b) and KIC 10403228: Over a single $\sim$56~day period J1407b was seen to undergo a series of deep and complex dimming events that were fitted with a comparatively complex set of no less than 37 separate rings \citep{Mamajek12, Werkhoven14}. More recently the single asymmetric transit-like event of KIC 10403228 was modelled as the transit of a grazing planet with an oblique ring \citep{Aizawa17}. Both of these cases are single events and lack confirmation from follow-up observations. In addition, \citet{Santos15} attempted to detect rings using the reflected light from the planet 51 Pegasi b. Despite the rings providing an adequate explanation for the reflected light, dynamical (and other) considerations make that explanation unlikely.

    Past work used mostly numerical approaches to modelling rings. In this paper, we present the Polygon+Segments (hereafter P+S) algorithm: an approach which can be used to efficiently model ringed exoplanets, without sacrificing accuracy. This approach can also be used to model oblate planets and other, more complex, configurations. We present our approach by gradually increasing the level of geometrical complexity: in \S \ref{UniformSource} we model the light curve of a transiting opaque circle, and then an opaque ellipse, before combining them to produce a model of a ringed planet transiting a uniform source. In \S \ref{NonUniform} we extend the above to non-uniform sources. In \S \ref{Comparison} we validate the results obtained using the Polygon+Segments model with numerical and other models discussed in the literature, and conclude in \S \ref{Conclusions}.

	\section{Uniform Source}
	\label{UniformSource}
	
	The analysis of the unphysical case of a completely uniform source has two advantages: firstly, its relative simplicity allows drawing conclusions regarding the relationships between the different parameters of the problem. Secondly, it turns out to be a key step towards the more general case of radially-symmetric sources discussed in \S \ref{NonUniform}.
	
	In the usual normalised units (un-obscured stellar flux is unity, stellar radius is unity) the flux during transit of a uniform source $f_\textrm{u}$ can be calculated directly from the hidden area by the transiting object:
	\begin{equation}
    	\label{eq:LC_uniform}
    	f_\textrm{u}=1-\frac{A_\textrm{h}}{A_\textrm{star}}= 1-\frac{A_\textrm{h}}{\pi}
	\end{equation}
	where $A_\textrm{h}$ is the hidden area. This allows stating the problem in geometrical terms of finding the overlap between two 2D shapes. Firstly, we will address the well-known case of a spherical planet, which appears as a circle to the observer, but in the context of the Polygon+Segments model. Then, we will add a layer of complexity by allowing the planet to be oblate, i.e. have a projected shape of an ellipse - which is useful since this is also the sky-projected shape of a disc. These two basic light curves cannot be simply added together to form the light curve of a ringed planet since some area will be hidden by both the planet and disc simultaneously, causing simple addition to double-count this 'doubly-hidden area' (hereafter DHA). We will therefore present an algorithm for the calculation of the instantaneous DHA. In practice, in order to account for finite integration times \citep{Kipping10} one may need to oversample and then integrate the instantaneous model presented here.
	
	\subsection{Circular planets}
    	The standard model for transiting exoplanets \citep[hereafter MA02]{MA2002} is that of a spherical planet which has a sky-projected shape of a circle. We introduce the Polygon+Segments algorithm in this well-known context.	We use a coordinate system centred on the planet (and not the star), such that points on the edge of the star satisfy: $(x-x_\textrm{s})^2+(y-y_\textrm{s})^2=1$ and points on the edge of the planet satisfy: $x^2+y^2=r_\textrm{p}^2$, where $(x_\textrm{s}, y_\textrm{s})$ is the centre of the star, and $r_\textrm{p}$ is the radius of the planet relative to its host. 
    	The curves of the planet and the star can have either zero, one or two intersection points. If there are no intersections, one curve is either completely inside the other or the curves are completely separate. If the star and the planet are tangent there is exactly one common point between them but this does not change the intersection area and the same logic as in the no-intersections case applies.
    	
    	\begin{figure}
    		\includegraphics[width=\linewidth]{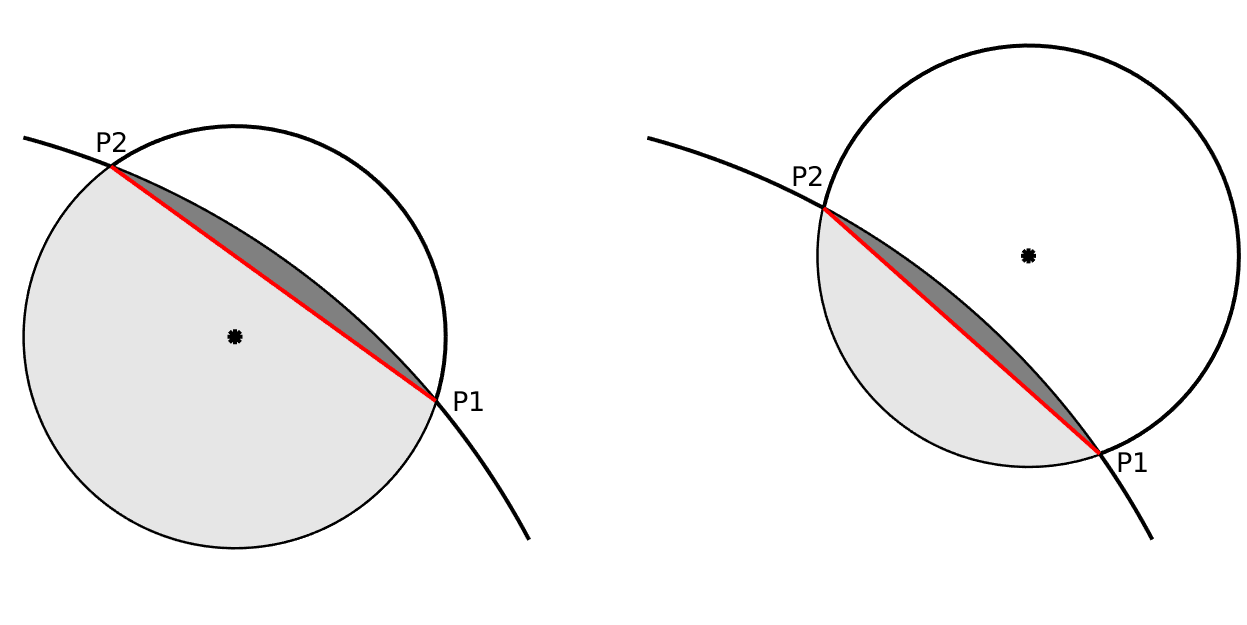}
    	    \caption{
    	    Examples for hidden area determination by splitting it into segments. The circular planet and a partial arc of the larger star are visible. The hidden area is the sum of a star segment (dark grey) and a planet segment (light grey), separated by a chord (red line). It does not matter if the planet's centre is inside (left) or outside (right) of the star's disc.}
    	    \label{fig:phasegments}
    	\end{figure}
    	
    	If there are two intersection points (see Figure \ref{fig:phasegments}), their location can be calculated from simple geometry \citep[e.g.][]{Bourke97}. Let us designate these two intersection points as $P_1=(x_1,y_1)$, and $P_2=(x_2,y_2)$. By construction, on each side of the chord connecting $P_1$ and $P_2$ there are two sectors -- one of the planet and one of the star. Since we are interested in the intersection between the two objects, the overlap on each side is just the smaller of the two sectors on that side of the chord, and the total hidden area is then the sum of these two smaller sectors:

        \begin{equation}
        	A_\textrm{h}=\left\{
        	{
        	\begin{array}{ll}
        	\min(S_\mathrm{P_1,P_2,Planet}, S_\mathrm{P_1,P_2,Star}) + \\
        	\quad \min(S_\mathrm{P_2,P_1,Planet}, S_\mathrm{P_2,P_1,Star})	&;  (1-r_\textrm{p})^2<x_\textrm{s}^2+y_\textrm{s}^2<(1+r_\textrm{p})^2 \\
        	\pi \min(r^2_\textrm{p},1) &;  x_\textrm{s}^2+y_\textrm{s}^2\le(1-r_\textrm{p})^2\\
        	0 &; \text{otherwise}
        	\end{array}
        	}
        	\right.
    	    \label{eq:pha}
    	\end{equation}
        
        where $S_\textrm{A,B,Curve}$ is the area of a segment from point A to B, going counterclockwise about the middle of the chord $\overline{AB}$ on Curve, and where Curve can be either the star or the planet (and later - the curves of the ring's inner or outer radii). In most physical arrangements, $r_\textrm{p}<1$, and thus $min(r_\textrm{p}^2,1)=r_\textrm{p}^2$. However, it will later be shown that accounting for limb-darkening requires us to allow $r_\textrm{p}>1$. Moreover, some physical host stars (e.g., white dwarfs, neutron stars) may be smaller than their planets. Thus, the geometry underlying this calculation does not rely on any assumption as to the relative radii of the planet. 

    \subsection{Oblate planets or inclined discs}	
    \label{UniformOblate}
        The logical procedure above remains even if the obstructing shape is an ellipse instead of a circle. The ellipse is assumed to be just a circle of radius $r_{ \rm e}$ viewed at an inclination angle $i$ relative to the line of sight. Therefore, the star's equation is unchanged at $(x-x_{ \rm s})^2+(y-y_{ \rm s})^2=1$, but the ellipse's equation is now $x^2+{y^2}/{\cos^2(i)}=r_{ \rm e}^2$. 
        Unfortunately, the oblate case lacks a simple relation to check whether there are intersections between the curves. Therefore, we calculate the intersections by finding the roots of a quartic polynomial, of which there are always four in complex numbers. Of these roots, each real root is associated with an intersection point, and when the polynomial has no real roots there are no intersection points.
        We note that although quartic equations are generally solvable, the solution is numerically unstable. We use \citet{Strobach10} to solve it efficiently and to high precision (close to machine precision, see also appendix \ref{SimilarRoots}).
    	
\begin{figure}
    		\includegraphics[width=\linewidth]{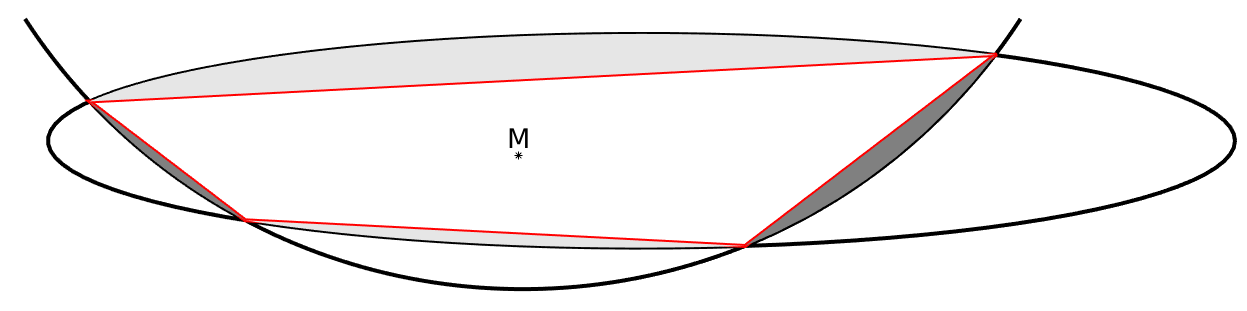}
    		\caption{An object's hidden area calculation - an example using a highly oblate object. The hidden area $A_\textrm{h}$ is split by chords to a polygon and segments of either the planet (light grey) or the star (dark grey). Note that by construction, between every two intersections there are segments of both the star and the planet, and the smaller of each pair is added to $A_\textrm{h}$.}
    		\label{fig:oblatealgex}
    	\end{figure}

        In cases where there are zero or one intersections, the calculation is similar to the circular case. If there are more intersections, one can express the total hidden area as the sum of a few basic shapes. Since the hidden area is an intersection of a circle and an ellipse, the set of lines azimuthally connecting all the intersection points would be a polygon with edges that are all chords of both the circle and the ellipse. The polygon would also be completely inside the hidden area, as seen in Figure \ref{fig:oblatealgex}. The polygon's area can be calculated without additional difficulties (see Equation \ref{eq:polyarea} below), as well as a point inside it $M$ (the mean of the coordinates of the vertices). At this point the rest of the hidden area can be found in a manner similar to the circular case: on the outer side of each chord (outside of the polygon) there are two segments and we seek to find their overlap, which again is simply the smaller of the two. The total hidden area is therefore the sum of the polygon and the smaller of each pair of outer segments between each pair of adjacent vertices. The adjacent vertices are found by ordering all vertices azimuthally about $M$. This guarantees both that adjacent vertices are identified as such, and that only the segments completely outside of the polygon are selected.

    	For convenience, we remind the reader of a few geometrical properties. The area of a convex polygon with vertices, $(x_1 ,y_1),(x_2,y_2),...,(x_\textrm{n},y_\textrm{n})$, sorted counterclockwise, is:
    	\begin{equation}
    	    A_{\rm polygon}=(x_1y_2-x_2y_1+x_2y_3-x_3y_2+...+x_\textrm{n-1}y_\textrm{n}-x_\textrm{n}y_\textrm{n-1}+x_\textrm{n}y_1-x_1y_\textrm{n})/2
    	    \label{eq:polyarea}
    	\end{equation}
    	In an ellipse with semi-major axis $a_{\rm e}$ and semi-minor axis $b_{\rm e}$ the area of a sector, as measured from the semi-major axis and relative to the ellipse's centre is:
    	\begin{equation}
    	    A_{\rm sector}(0,\beta)=
                ({a_{\rm e}b_{\rm e}[\frac{\beta}{\pi}]\pi} + {a_{\rm e}b_{\rm e}\arctan(\frac{a_{\rm e}}{b_{\rm e}}\tan(\beta-\pi [\frac{\beta}{\pi}]))})/2 
            \label{eq:secarea}
    	\end{equation}
    	where $[x]$ is the round operation of $x$ to the nearest integer and $A_{\rm sector}(0,\beta)$ is the area of a sector from the semi-major axis whose central angle is $\beta$. From Equation \ref{eq:secarea} it follows that the area of a sector between angles $\beta_1$, $\beta_2$ (where $\beta_1\leq\beta_2$) is  $A_{\rm sector}(\beta_1,\beta_2)=A_{\rm sector}(0,\beta_2)-A_{\rm sector}(0,\beta_1)$. Lastly, for a circle, the area of a sector with central angle $\beta$ reduces to $A_{\rm sector}(\beta)={\beta}r^2/2$.

	\subsection{Exoplanets with rings}
	\label{subsec:exorings}
	We model a circumplanetary ring $R$ as the area between two concentric discs, $D_\textrm{in}\subset D_\textrm{out}$, viewed at the same inclination angle. In that area, only a portion of the light, the ring's opacity $0<w<1$, is blocked. We express the total effective hidden area $A_\textrm{h}$ as a sum of terms, each is an intersection of a different subset of the problem's convex shapes (e.g., circles, ellipses, etc.). In our case, if we mark the planet $P$ and the star $S$, we obtain that the total effective hidden area is:
    \begin{equation}
        \begin{array}{rl}
        A_\mathrm{h} & =A_\mathrm{P\cap S}+w*(A_\mathrm{R\cap S}-A_\mathrm{P\cap R\cap S}) \\
            & =A_\mathrm{P\cap S}+w*(A_\mathrm{D_{out}\cap S}-A_\mathrm{D_{in}\cap S}-A_\mathrm{P\cap D_{out}\cap S}+A_\mathrm{P\cap D_{in}\cap S}) \\
        \end{array}
        \label{eq:ringuni}
    \end{equation}
    where $A_\mathrm{C_i \cap C_j}$ for any curve C is the intersection area between the curves $C_{\rm i},C_{\rm j}$. The problem is now reduced to finding the area hidden by multiple curves simultaneously. For this purpose, we introduce the Polygon+Segments algorithm (P+S) which builds on the simpler problems of the previous sections. 
    
	\noindent  \fbox{
 \ \ \begin{minipage}{.9\linewidth}
		\textbf{Polygon+Segments Algorithm}\\
			Goal: calculation of the intersection area, $A_{\rm double}$, of all convex shapes defined by curves $C=\left\{C_1,C_2,...,C_\textrm{k}\right\}$:\\
			\begin{legal}
    		\item Determine all intersection points ${\hat{P}}=\left\{P_1,P_2,...,P_\textrm{m}\right\}$ of any two curves $\in C$. From which:
    		\begin{legal}
        		\item Define subset $P$ of ${\hat{P}}$ which are the points on the border $A_{\rm double}$. Subset
        		$P=\left\{P_1,P_2,...,P_\textrm{n}\right\}$ is defined by: points $\in {\hat{P}}$ that are in or on all curves. 

        		    If P is not empty - continue to step 2. If P is empty at least one of the curves is disjoint. If there is a curve $C_\textrm{j}$ completely inside all other curves, \textbf{return} $A_\textrm{double}\leftarrow C_\textrm{j}$'s area and exit. Otherwise, \textbf{return} $A_\textrm{double}\leftarrow 0$ and exit.
        		\end{legal}
        		\label{step:points_test}
                \item Determine the coordinates of $M$, the average of the coordinates of the intersection points $\in P$.
                Sort $\left\{P\right\}$ azimuthally relative to $M$. Note that this means that the point following $P_{\rm j}$ is known: $P_\textrm{next}\equiv P_{\rm j+1\mod\,n}$
                \item Calculate initial hidden area: $A_{\rm double} \leftarrow$ area of the polygon defined by $\left\{P\right\}$.
    		\item \textbf{For} intersection point $P_{\rm j}$. \textbf{Step:}
    		\begin{legal}
        		\item For each curve $\in C$ that $P_{\rm j}, P_\textrm{next}$ are both on, calculate the curve's segment area between these two points.
        		\label{step:curve_test}
        		\item Add the minimum of all such areas to $A_{\rm double}$.
    		\end{legal}
    		
    		\item \textbf{return} $A_{\rm double}$
    		\end{legal}
		\end{minipage}
    }
    
    In the case of two curves ($k=2$), all of the intersection points are on all curves so steps \ref{step:points_test} and \ref{step:curve_test} can be simplified. Note that some subtle numerical issues are addressed in Appendix \ref{ToleranceParameter}.
    
\begin{figure}
		\includegraphics[width=0.9\linewidth]{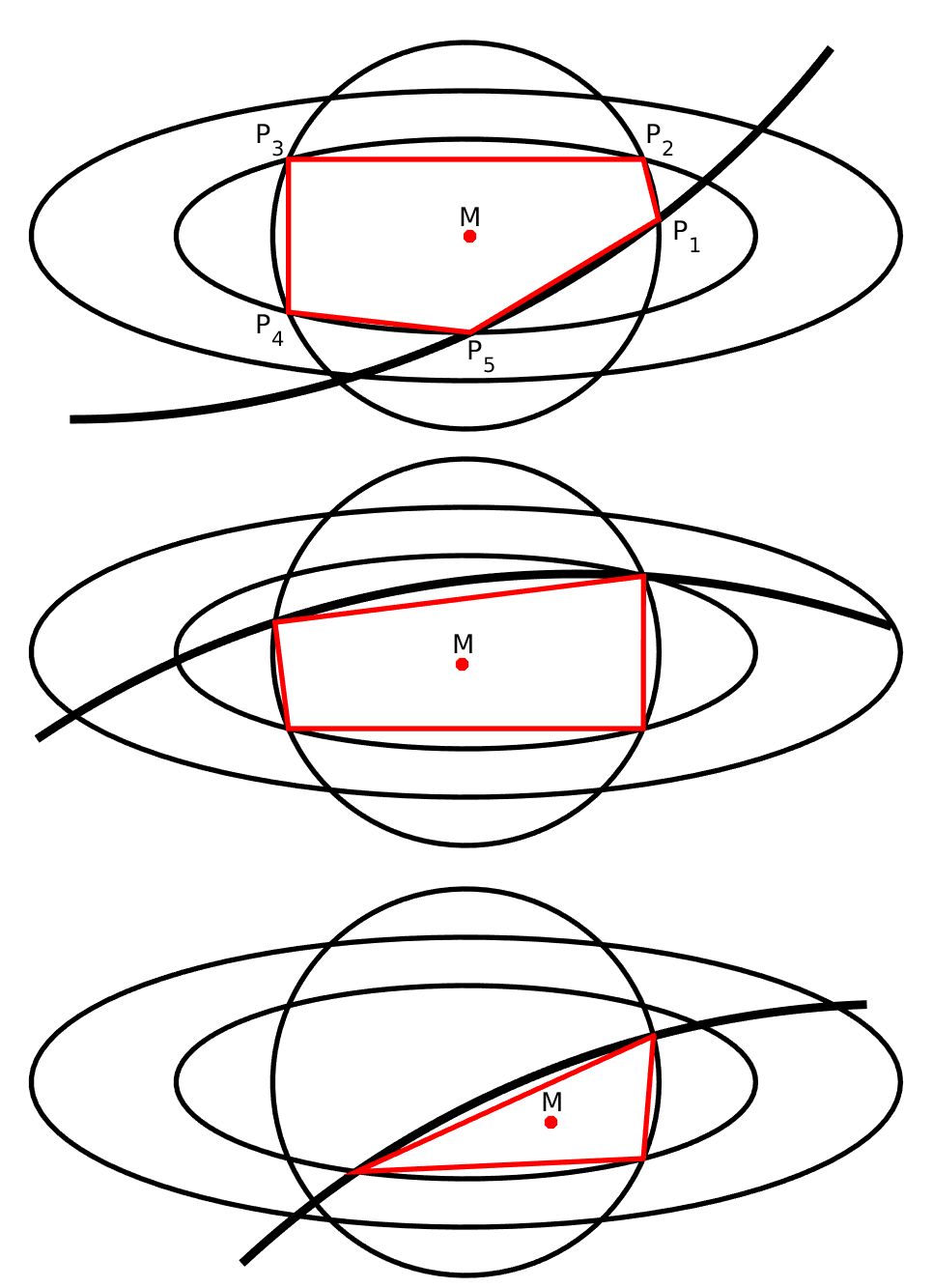}
		\centering
		\caption{Doubly hidden area (DHA) polygon and segment division examples. All panels show the calculation of $A_\mathrm{P\cap D_{in}\cap S}$ (which is one term in eq. \ref{eq:ringuni}, also expanded upon in the text). The thick curve is part of the star, the circle is the planet and the ellipses are the ring's edges. The DHA is divided by thin red lines (which are a set of chords of the above shapes) into a polygon and several segments. The point $M$ is the centre of the polygon. \textbf{Top:} Points $P_1$ through $P_5$ are the intersection points on the border of the DHA (step \ref{step:points_test}), counterclockwise-ordered with respect to $M$. On each step the area of all the outer segments subtended between the start \& end intersections defining that segment are calculated, and the smallest one is added to the DHA. \textbf{Middle:} similar to the above, but for a configuration that includes a triple intersection. \textbf{Bottom:} similar to the above, now with the planet's centre off the star, to emphasise that $M$ is not the centre of the planet.
		}
		\label{fig:dha}
	\end{figure}

	Since this is rather abstract, we explicitly write below and plot some of the steps as an example: each of the terms in eq. \ref{eq:ringuni} is calculated using P+S. In our problem, there are four primary curves of S, P, $\mathrm{R_{in}}, \mathrm{R_{out}}$ for the star, planets and ring radii. In the geometry shown in the top panel of Figure \ref{fig:dha} the last term is calculated in the following manner:

    \begin{equation}
        \begin{array}{l}
        A_\mathrm{P\cap D_{in}\cap S} = \\
        \begin{array}{lll}
            1.  & & \hat{P}=\{\mathrm{All\,\,intersections\,\,of\,\,any\,\,pair\,\,of:\,\,P,}\,\,\mathrm{D_{in}}\,\,\mathrm{and\,\,S}\} \\
                & 1.1.  & P=\{P_1, P_2, P_3, P_4, P_5\} \\
            2.  & & \mathrm{Define}\,\, M \\
            3.  & &    A_\textrm{h}= \mathrm{area\,\,of\,\,polygon}\,\,P \\
            4   & 4.1 &   (P_j,P_{next})=(P_1,P_2) \\
                &     & \rotatebox[origin=c]{180}{$\Lsh$}  \mathrm{Current\,\,curve} = \mathrm{P} \\
                & 4.2 & \rotatebox[origin=c]{180}{$\Lsh$} \,\,A_h = A_h + \mathrm{P\,\,segment\,\,between\,\,}P_1\,\,\mathrm{and\,\,}P_2 \\
                & 4.1 &    (P_j,P_{next})=(P_2,P_3) \\
                &     & \rotatebox[origin=c]{180}{$\Lsh$}  \mathrm{Current\,\,curves} = \mathrm{D_{in},\,\,P} \\
                & 4.2 & \rotatebox[origin=c]{180}{$\Lsh$} \,\,A_h = A_h + \mathrm{D_{in}\,\,segment\,\,between\,\,}P_2\,\,\mathrm{and\,\,}P_3 \\
                & 4.1 &    (P_j,P_{next})=(P_3,P_4) \\
                &     & \rotatebox[origin=c]{180}{$\Lsh$}  \mathrm{Current\,\,curves} = \mathrm{D_{in},\,\,P} \\
                & 4.2 & \rotatebox[origin=c]{180}{$\Lsh$} \,\,A_h = A_h + \mathrm{P\,\,segment\,\,between\,\,}P_3\,\,\mathrm{and\,\,}P_4 \\
                & 4.1 &    (P_j,P_{next})=(P_4,P_5) \\
                &     & \rotatebox[origin=c]{180}{$\Lsh$}  \mathrm{Current\,\,curve} = \mathrm{D_{in}} \\
                & 4.2 & \rotatebox[origin=c]{180}{$\Lsh$} \,\,A_h = A_h + \mathrm{D_{in}\,\,segment\,\,between\,\,}P_4\,\,\mathrm{and\,\,}P_5 \\
                & 4.1 &   (P_j,P_{next})=(P_5,P_1) \\
                &     & \rotatebox[origin=c]{180}{$\Lsh$}  \mathrm{Current\,\,curve} = \mathrm{S} \\
                & 4.2 & \rotatebox[origin=c]{180}{$\Lsh$} \,\,A_h = A_h + \mathrm{S\,\,segment\,\,between\,\,}P_5\,\,\mathrm{and\,\,}P_1 \\
        \end{array}        
        \end{array}
\end{equation}
	
	Thus, we can model light curves of planets with rings transiting a uniform star, and indeed of any object which is a combination of convex primitive shapes.
	
    We note that equation \ref{eq:ringuni} is a special case of, and can be generalised by, the inclusion-exclusion principle. Using it we can express the total hidden area of any set of curves using intersection areas only:
	\begin{equation}
	    A_\textrm{h}=|\bigcup\limits_{\rm j=1}^{n} C_{\rm j}| = \sum\limits_{k=1}^{n} (-1)^{k+1} \sum\limits_{1\leq \rm j_1<\rm j_2<...<\rm j_k\leq n} |C_{\rm j_1}\cap C_{\rm j_2} \cap ... \cap C_{\rm j_k}|
	    \label{eq:includeexclude}
	\end{equation}
	where $C_{\rm j}$ is the j-th curve, and for convex curves each of these terms can be calculated by the P+S algorithm. The corresponding light curve is found using $A_\textrm{h}$ and equation \ref{eq:LC_uniform}. 
	
	\section{Non-Uniform Source}
	\label{NonUniform}
	
	In practice, stars do not have uniform surface brightness and exhibit limb darkening and other non-uniformities. Thus, the stellar local intensity is commonly expressed as a function of a point's distance from the star's centre $r$. 
	A useful connection between the uniform and non-uniform cases is given in MA02 which we reproduce here:
	\begin{equation}
    	F_{\rm MA}(r_{\rm p},z)=\frac{\int_{0}^{1}drI(r)\frac{d[F_{\rm MA}^e(r_{\rm p}/r,z/r)r^2]}{dr}}{\int_{0}^{1}dr2rI(r)}
	    \label{eq:ma}
	\end{equation}
	
	where $r_{\rm p}$ is the normalised planet's radius ($p$ in MA02), $z$ is the normalised sky-projected planet-star distance, $I(r)$ is the local intensity of the star, and $F_{\rm MA}^{e}(r_{\rm p},z)$ is the uniform source model. Note that the division of the size $r_p$ and distance $z$ by $r$ means that the accompanying uniform source model must be evaluated also when $r_\textrm{p}>1$. This can be extended to oblate planets and planets with rings since this equality does not depend on the radial symmetry of the planet. By simply changing the arguments of $F^{e}$ one obtains for oblate planets:
	\begin{equation}
    	F_{\rm oblate}(a,b,x_\textrm{p},y_\textrm{p},\theta,i)=\frac{\int_{0}^{1}drI(r)\frac{d[F_{\rm oblate}^e(r_{\rm e}/r,x_\textrm{p}/r,y_\textrm{p}/r,\theta,i)r^2]}{dr}}{\int_{0}^{1}dr2rI(r)}
    	\label{eq:ma_oblate}
	\end{equation}
	where $r_\textrm{e}$ is the oblate planet's sky-projected semi-major axis, $(x_{\rm p},y_{\rm p})$ are the planet's centre's coordinates, $\theta$ the projected obliquity angle of the planet, $\cos(i)$ is the ratio between the planets' minor and major axes, parameterised as an angle as in \ref{UniformOblate}, and $F_{\rm oblate}^{e}$ is the accompanying uniform source model. 
	
	Similarly, for planets with rings:
	\begin{equation}
        \begin{array}{l}
	        F_{\rm rings}(r_\textrm{p},r_{\rm in},r_{\rm out},x_p,y_p,i,\theta,w)= \\
	        \quad \quad \frac{\int_{0}^{1}drI(r)\frac{d[F_{\rm rings}^e(r_\textrm{p}/r,r_{\rm in}/r,r_{\rm out}/r,x_p/r,y_p/r,i,\theta,w)r^2]}{dr}}{\int_{0}^{1}dr2rI(r)}
        \end{array}
    \label{eq:ma_ringed}
	\end{equation}
	where $r_{\rm p}$ is the planet's radius, $r_{\rm in}$ is the inner radius of the ring, $r_{\rm out}$ is the outer radius of the ring, $(x_{\rm p},y_{\rm p})$ are the planet's centre's coordinates, $\theta$ is the projected obliquity angle of the ring, $i$ the ring's inclination, $w$ is the ring's opacity (see \ref{subsec:exorings}) and $F_{\rm rings}^{e}$ is the accompanying uniform source model. Note that the non-length variables $\theta,i,w$ are not divided by $r$ in the integral.	\\
	
	We can simplify the expressions for equations \ref{eq:ma}, \ref{eq:ma_oblate} and \ref{eq:ma_ringed} using integration by parts:
	\begin{equation}
	F(\boldsymbol{\mathit{d}},\boldsymbol{\mathit{\alpha}})=\frac{I(1)F^e(\boldsymbol{\mathit{d}},\boldsymbol{\mathit{\alpha}})-\int_{0}^{1}dr\frac{dI(r)}{dr}F^e(\boldsymbol{\mathit{d}}/r,\boldsymbol{\mathit{\alpha}})r^2}
	{\int_{0}^{1}dr2rI(r)}
	\label{eq:intparts}
	\end{equation}
	where $\boldsymbol{\mathit{d}}$ represents input parameters that are related to length: sizes and distances, and $\boldsymbol{\mathit{\alpha}}$ stands for non-length parameters: angles and the opacity.
	The intensity function, i.e. the limb-darkening law, is analytic and known in advance so finding its derivative is simple. \\
	
	One of the most generalised limb darkening (LD) expressions is the four-parameter, non-linear LD law $I(r)=1-\sum_{n=1}^{4} c_\textrm{n}(1-(1-r^2)^{n/4})$ \citep[see][]{Claret13}. Its derivative is $I'(r)=-r\sum_{n=1}^{4} \frac{n}{2}c_\textrm{n}(1-r^2)^{(n-4)/4}$. Usually, these LD laws are not written as a function of $r$ but rather of $\cos{\phi}=\sqrt{(1-r^2)}$, where $\phi$ is the angle between the stellar surface and the line of sight. 

	Substituting $t\equiv\sqrt{\cos{\phi}}=(1-r^2)^\frac{1}{4}$ we obtain:
	\begin{equation}
	    \begin{split}
             \int_{0}^{1}dr\frac{dI(r)}{dr}F^e(\boldsymbol{\mathit{d}}/r,\boldsymbol{\mathit{\alpha}})r^2 =\\
	          -\int_{0}^{1}dt
                \sum_{n=1}^{4} nc_nt^{n-1}
                F^e(\frac{\boldsymbol{\mathit{d}}}{\sqrt{1-t^4}},\boldsymbol{\mathit{\alpha}})(1-t^4)
	    \end{split}
	\end{equation}
	
	The integrand can accommodate either the quadratic, three-parameter \citep{Kipping16} or the non-linear LD laws. Using this substitution, most commonly used limb-darkening laws can be used without difficulty or performance issues. 
	Also, using the model introduced by \citet{Claret18}, which introduced a critical angle $\mu_\textrm{crit}$, only requires that the integral over $r$ be evaluated not on $[0,1]$ but rather on $[0,\sqrt{1-\mu_\textrm{crit}^2}]$,
	and after the substitution above, it becomes $[\sqrt{\mu_\textrm{crit}},1]$. The same change is needed in the limits of the integrals of eq. \ref{eq:intparts}.  See Appendix \ref{NumericalIntegration} for details on the numerical integration.

    In some cases, it may be more efficient to use table interpolation instead of direct calculation of model values. For example, when fitting a model to some data, one often calls the model function many times with similar input parameters. One may choose to calculated a large table \textit{a priori} using the P+S algorithm, covering all geometries of a given star, and interpolate the model values from that table during the fitting itself. This approach is highly parallelisable, and it changes the computational burden from a cumulative one, which grows as additional calls are made to the model function, to one where most costs are incurred initially. For the tests conducted in this paper, we use the model directly, without the use of a table contemplated above.

	\section{Model Validation and Comparison}
	\label{Comparison}

	\textbf{Model validation:} we test the implementation of the P+S model, the Python code package \texttt{pyPplusS}, in order to validate its results.
	
	\begin{enumerate}
	    \item Ringless spherical planet: As a first test, we compared our model with the standard MA02 model. Simulating a typical hot Jupiter, the models agree to high accuracy ($O(10^{-6})$, see Figure \ref{fig:MAcomp}). Differences are always far smaller than typical Kepler errors, but peak near $|z|=1-r_\textrm{p}$. On the one hand, the MA02 model itself is less precise at this point \citep[see][Section 3.5]{Ofir18}. On the other hand, the models' differences become smaller as the integration order is increased. We therefore cannot easily assign an origin to this discrepancy. We repeated this check in a grid covering the most relevant parameter values: planet radii in the range [0.01, 0.25] and impact parameters in the range [0, 1.2]. In all cases the maximal error was $\sim10^{-5}$ and $\sim5.8\cdot10^{-7}$, for n=5 and n=10 integration orders, respectively. We find that the differences are always far smaller than typical Kepler errors.

    	\begin{figure}
    		\includegraphics[width=\textwidth]{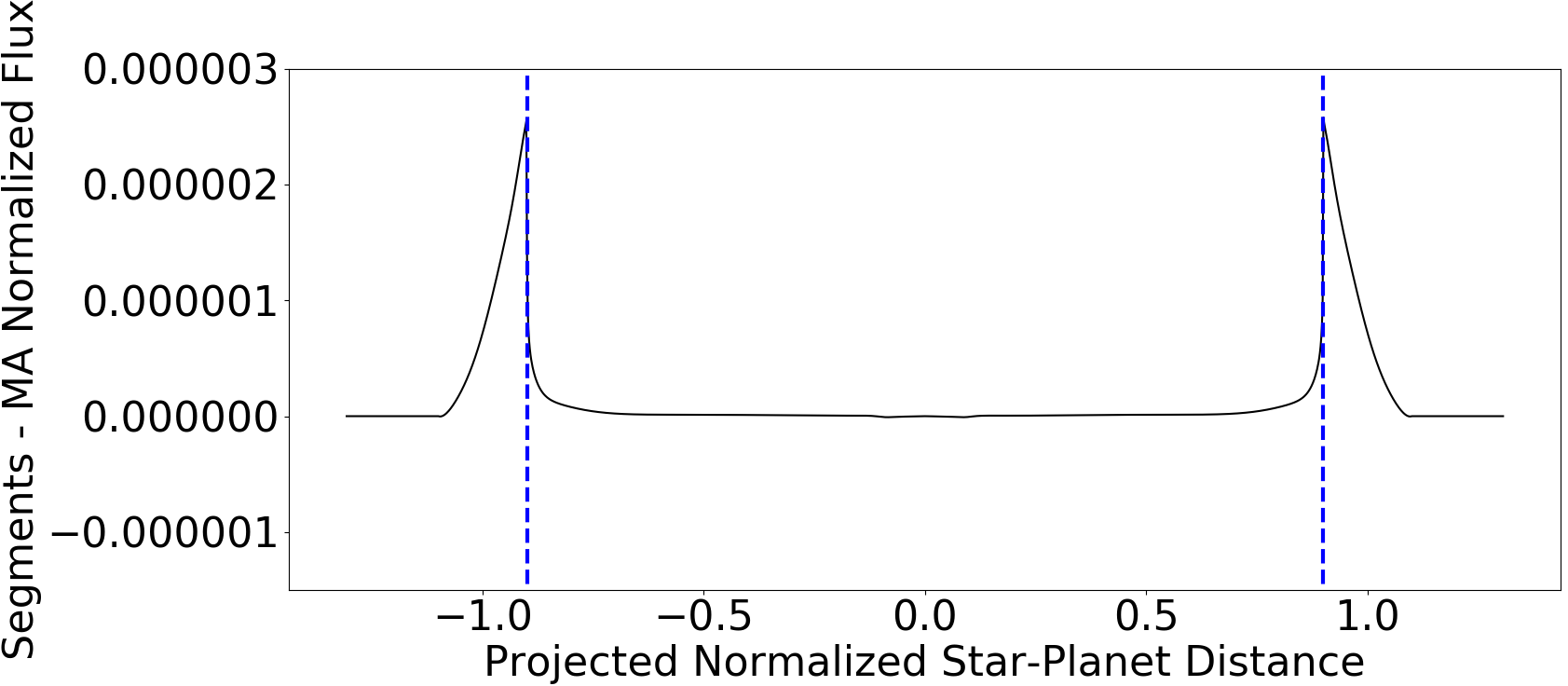}
    		\caption{The difference between the MA02 model and the \texttt{pyPplusS} model for a centrally-transiting ($b=0$) hot Jupiter. The differences between the models are always significantly smaller than the errors in \textit{Kepler} observation, and peak near $|z|=1-r_\textrm{p}$ (vertical dashed lines). See discussion in the text for the origin of these small differences.}
    		\label{fig:MAcomp}
    	\end{figure}
	
        \item Oblate Planets: We describe an oblate planet using the P+S model as an opaque ring without a planet. In Figure \ref{fig:B&F6} we reproduce the results of \citet{BarnesFortney03} for the detectability of oblate planets (their Figure 6) using the \texttt{pyPplusS} and find that they are visually indistinguishable.

        \begin{figure}
    		\includegraphics[width=\textwidth]{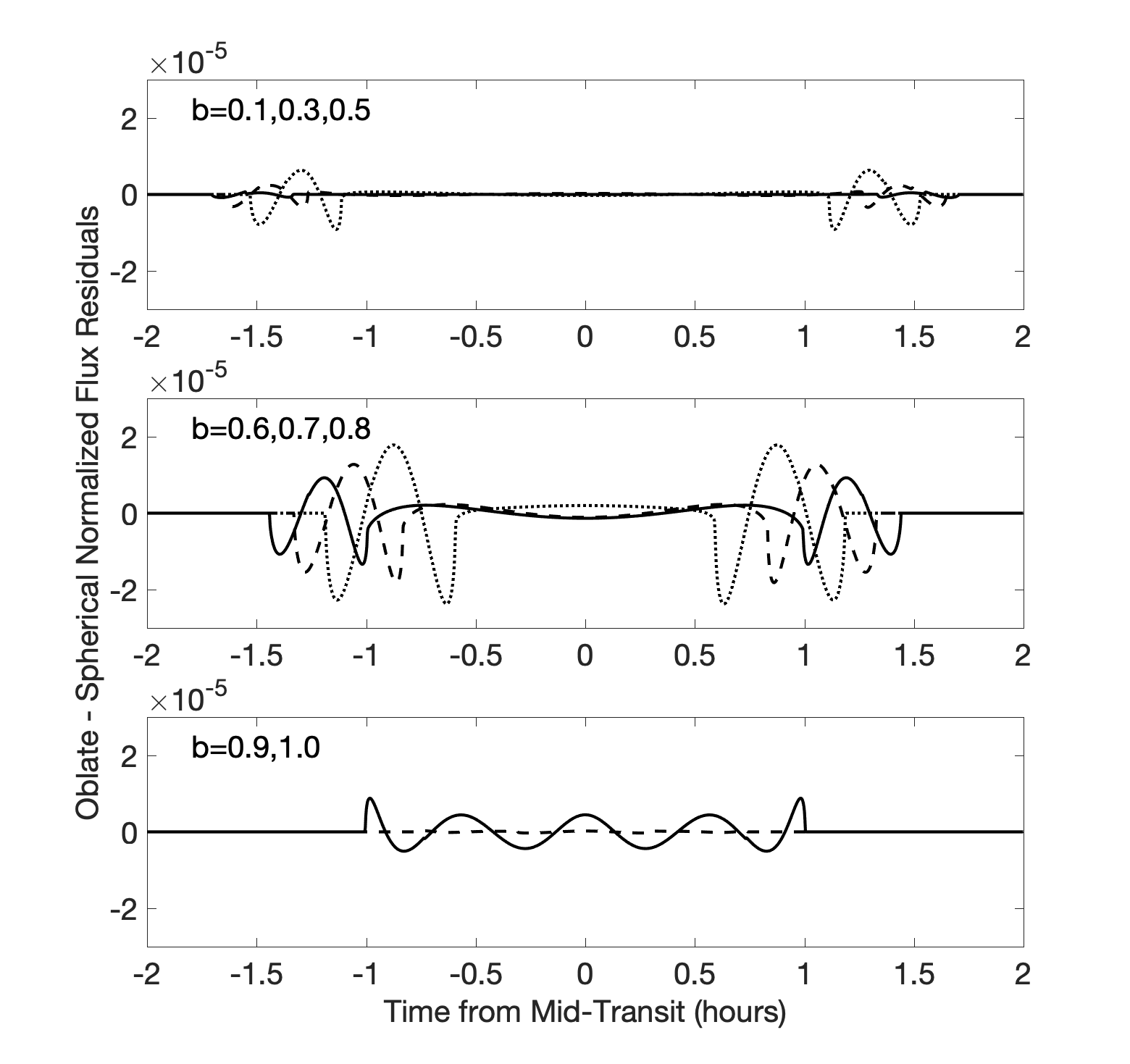}
    		\caption{A reproduction of figure 6 from \citet{BarnesFortney03} using the Polygon+Segments algorithm. These figures show the difference between an oblate planet signal and the best-fitting spherical model, allowing its radius, impact parameter $b$, the semi-major axis, and limb darkening coefficients to vary. \textbf{Top:} $b=0.1, 0.3,$ and $0.5$ using solid, dashed and dotted lines, respectively. \textbf{Middle:} $b=0.6, 0.7$ and $0.8$ using solid, dashed and dotted lines, respectively. \textbf{Bottom:} $b=0.9$ (solid line), $b=1.0$ (dashed line). The resulting curves are visually very similar to the ones given by \citet{BarnesFortney03}.}
    		\label{fig:B&F6}
        \end{figure}
        
        \item Ringed planets, numerical validation: we compare \texttt{pyPplusS} to two numerical models we developed. These numerical models estimate the light curve value by generating a discrete (pixelated) image of the star and the planet, with a finite resolution, and counting the flux of the pixels hidden by the ringed planet, similar to the technique used by \citet{Tusnski11,Akinsanmi18} and others. We developed two variants of the numerical models which we used for to validate \texttt{pyPplusS}: one in which the pixels are given on a uniformly-spaced grid ($Res$ elements in one stellar radius) and one in which the pixels are uniformly randomly distributed with an identical surface density of $Res^{-2}$. The former allows, in the case of a uniform source, to compare to exact results (e.g. a spherical planet with $r_\textrm{p}=0.1$ would produce a transit depth of exactly 1\%). Grid sampling, however, is prone to strong systematic errors close to sampling resonance. Random sampling is not susceptible to these errors and it can also provide an estimate for the numerical accuracy (via counting statistics of the stellar surface) -- but it can't provide exact results. We further improved the statistics of our numerical model by running it $N$ times and using the average of these runs and consequently scaling the standard deviation by $1/\sqrt{N}$ thus reducing the effect of any specific realisation of grid points. The expected numerical modelling error can be estimated in advance: The number of pixels in the star is just its area, i.e. about $N_S=\pi Res^2$, and this number is constant during the simulation. Similarly, the number of pixels in the planet, $N_P$, is roughly its area (regardless of shape), which for a circular planet of radius $r_p$ is $N_P=\pi (r_p Res)^2$. Note that $N_P$ is not constant due to pixel counting statistics as the planet moves across the face of the star, and we thus expect an uncertainty of $sqrt(N_P)$ (or $sqrt(\pi) r_p Res$ in the circular planet case). The relative error on the depth model will therefore be:
        \begin{equation}
            \sigma_\textrm{model}=\frac{sqrt(N_P)}{N_s} \quad \stackrel{\mathcal{\normalfont\mbox{circular}}}{=} \quad  \frac{r_\textrm{p}}{\sqrt{\pi N} Res}
            \label{eq:errest}
        \end{equation}
        
        This estimate will hold as long as the smallest feature on the (non-circular) planet will be resolved by $Res$. The \texttt{pyPplusS} model of a ringless planet agrees with both variants and a comparison between the random grid numerical variant and the \texttt{pyPplusS} model above is given in Figure \ref{fig:numcomp}. The agreement (within numerical error) of the models shows that the two models are consistent, and also agree with the expected error estimate. The error estimate for the ringed case requires the calculation of the ringed planet's hidden area. This calculation was done using the analytic \texttt{pyPplusS} algorithm.  We compared the numerical model and the \texttt{pyPplusS} algorithm for many parameter sets representing ringed exoplanets. In all tests, the models agreed within numerical errors. We do not plot the results of these other tests since they all but repeat Figure \ref{fig:numcomp}.
        Note that the error estimate, calculated using equation \ref{eq:errest}, was larger for the ringed exoplanet due to its larger hidden area. The constant resolution kept $N_s$ unchanged, while the larger planet (and addition of rings) caused $N_P$ to increase. Thus, the error estimate is larger.

    	\begin{figure}
    		\includegraphics[width=0.65\textwidth]{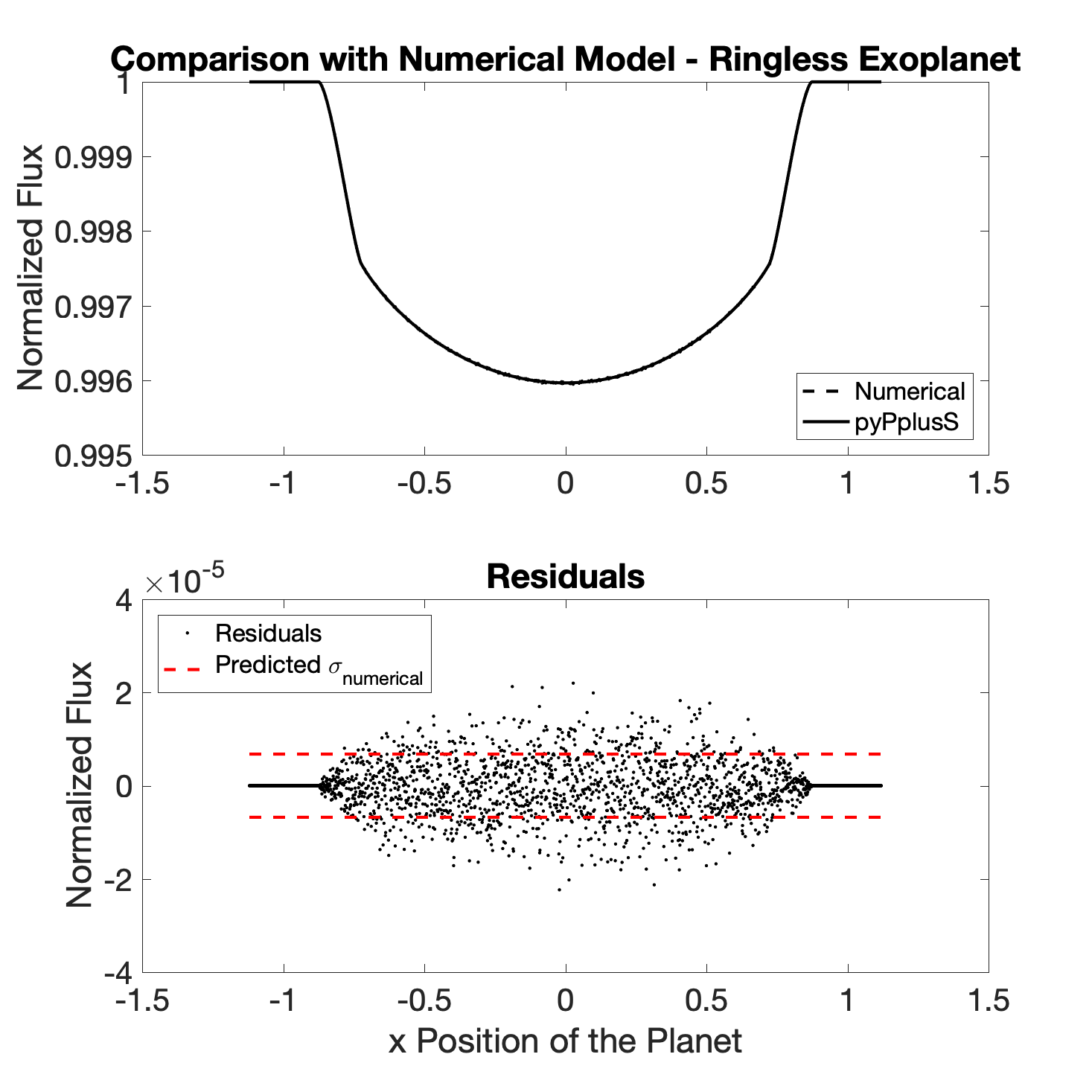}
    		\includegraphics[width=0.65\textwidth]{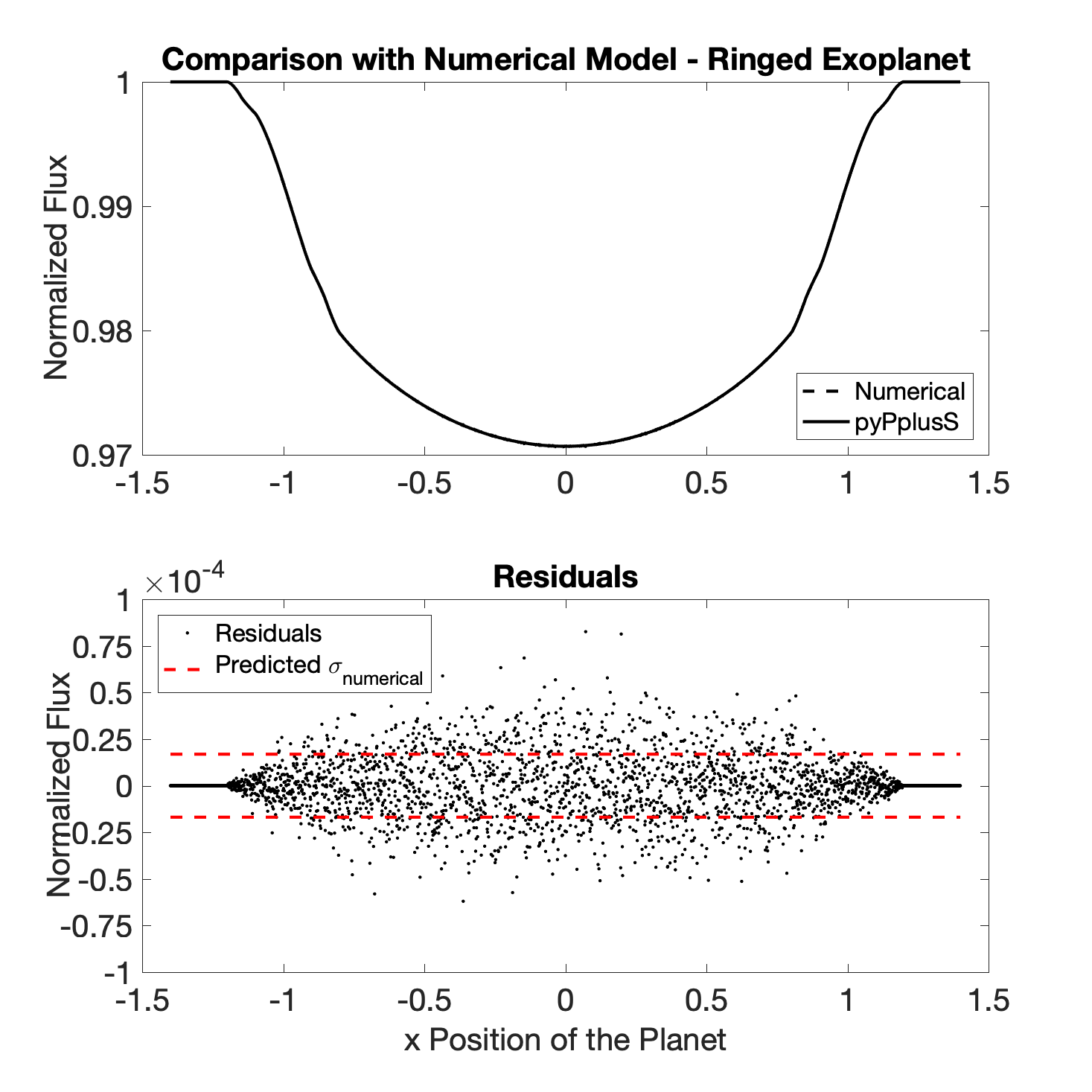}
    		\centering
    		\caption{A comparison between the numerical model, averaged over 25 runs, using $Res = 1000$ and the \texttt{pyPplusS} model in two configurations. \textbf{top:} a Saturn-sized, ringless exoplanet: $r_\textrm{p}/r_*=0.06, c_1=c_3=0, c_2=0.398667, c_4=0.263276, b=0.6$. \textbf{bottom:} a ringed exoplanet with an opaque ring: $r_\textrm{p}/r_*=0.1, r_{\rm in}/r_*=0.15, r_{\rm out}/r_*=0.2, i=45^{\circ}, \theta = 10^{\circ}, w=1.0, u_1=0.35, u_2=0.25, b=0$. \textbf{top panels:} In both configurations, the two model light curves are indistinguishable. \textbf{Bottom panels:} both panels show the difference between \texttt{pyPplusS} and the numerical model (note the y-axis scale). According to Equation \ref{eq:errest}, the numerical model is predicted to be precise to $\sigma_\textrm{model}\sim6.77\cdot 10^{-6}\,,\,\sim1.69\cdot 10^{-5}$ (top and bottom configurations, respectively, marked as red horizontal dashed lines) - and indeed the deviations are of that 
    		scale.} 
    		\label{fig:numcomp}
        \end{figure}

        \item Ringed planets, validation using literature codes: A few other ring-modelling codes were described in past literature (see model comparison below) - but only one was found to be publicly available: \texttt{exorings}\footnote{Available at https://github.com/mkenworthy/exorings}, developed by \citet{Kenworthy15}. We note that \texttt{exorings} was developed in the context of large rings with an apparent size much larger than the star. We found \texttt{exorings} performed poorly in the inverse case of a small ring (see below) and therefore tested \texttt{pyPplusS} at the \texttt{exorings} regime of a small star. We also found that \texttt{exorings} suffers from significant numerical instability: results at adjacent odd- and even- resolution values differ significantly (order of $10^{-2}$). Still, we compared the results of \texttt{pyPplusS} and \texttt{exorings} for the same parameter set representing a large and opaque ring, as described in Figure \ref{fig:kenworthy_comp}. This figure shows the small disagreements between the results can all be attributed to numerical errors (\texttt{exorings}' resolution). This is supported by the fact that the disagreement decreases as the number of pixels used to draw the star in \texttt{exorings} increases.
    	
    	\begin{figure}
    	    \includegraphics[width=\textwidth]{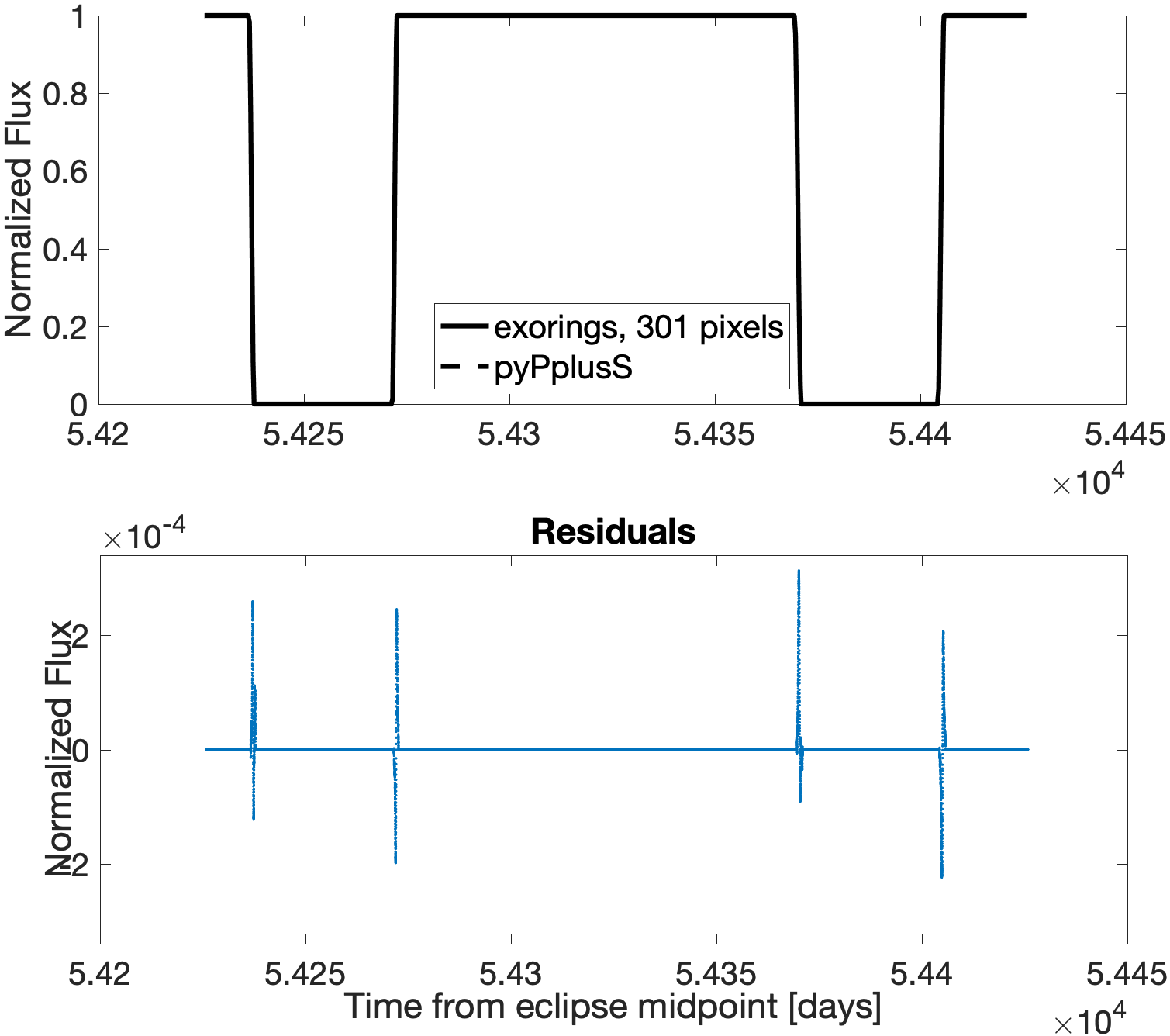}
    	    \caption{Comparison between \texttt{pyPplusS} and \texttt{exorings}. Note the geometry used by \texttt{exorings} is that of a very large ring with a small star. In this configuration, the star is eclipsed by the ring twice in a transit. In this comparison, $b=-9.218,r_{\rm in}=137.219,r_{\rm out}=232.575,i=69.454^{\circ},\theta=166.147^{\circ},u1=0.8$. We set $r_p=0.02$, as the planet is too small ($r_p<|b|-1$) to affect the light curve. The residuals occur at ingress and egress of the star behind the ring and with amplitude comparable to (\texttt{exorings}' resolution)$^{-2}$.}
    	    \label{fig:kenworthy_comp}
    	\end{figure}
    	
    	
    	In order to compare \texttt{pyPplusS} to \citet{BarnesFortney04}, we simulated a Saturn-like planet and fitted it with the MA02 model. For simpler discussion at this stage, we set the opacity to be unity, i.e. the ring doesn't allow any light to pass through it. The results can be seen in Figure \ref{fig:RingedBF}. This can be compared directly with \citet[fig. 1]{BarnesFortney04}, which is visually very similar - providing further validation of \texttt{pyPplusS}. By its nature, this figure also allows one to examine the detectability of a ringed exoplanet - as discussed in \citet{BarnesFortney04}.

    	\begin{figure}
    		\includegraphics[width=\textwidth]{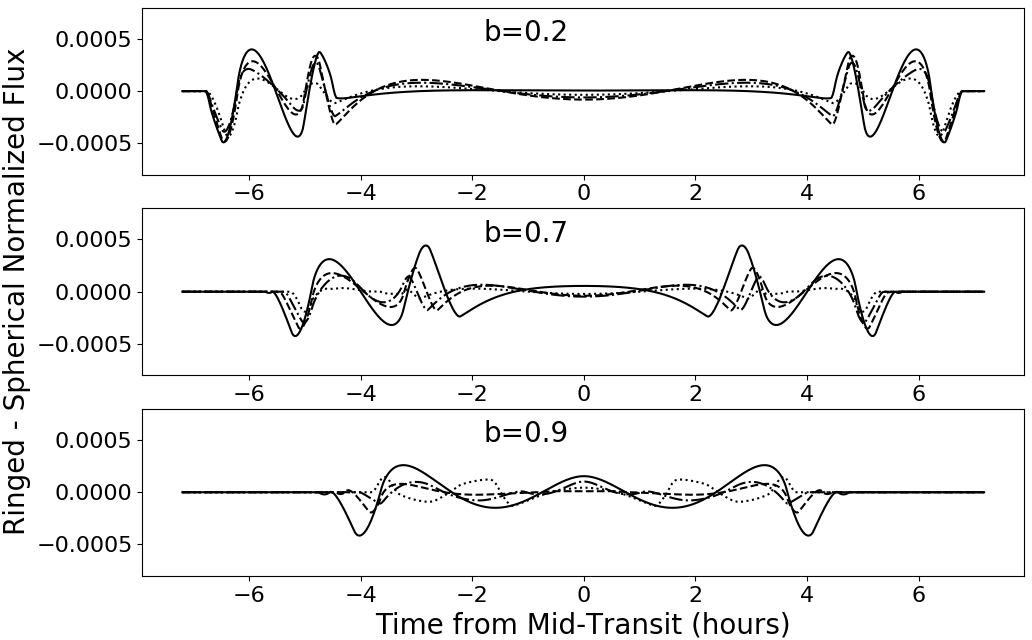}
    		\caption{The detectability of a ringed planet, i.e. the difference between a simulated ringed planet signal and the best-fitting ringless model, for a variety of impact parameters $b$. The star's limb darkness parameters are $u1 = 0.35; u2 = 0.25$ and the ringed exoplanet's parameters are $r_p=0.1, r_{in}=1.5*r_p, r_{out}=2.0*r_p, \theta=0, w=1-exp(-1/\cos(i))$ and the inclination is $0$ (face on), $45,60,80 $ for the solid, dashed, dot-dashed and dotted lines, respectively \textbf{Top:} $b=0.2$, \textbf{Middle:} $b=0.7$, \textbf{Bottom:} $b=0.9$}
    		\label{fig:RingedBF}
        \end{figure}
    \end{enumerate}
    
    We find that the \texttt{pyPplusS} code we presented here is consistent with our own numerical modelling, i.e. the pixelated code used for validation in Section \ref{Comparison}, Model validation, (iii), with the \texttt{exorings} code, as well as the results of \citet{BarnesFortney03, BarnesFortney04}. We conclude that \texttt{pyPplusS} is validated for correctness.
    
    \textbf{Model comparison:} Below we present literature relevant to \texttt{pyPplusS}. We compare \texttt{pyPplusS} with these studies with respect to their capabilities and/or performance. Some of them are directly comparable to \texttt{pyPplusS} in that they are supposed to allow the detection of ringed exoplanets, but most only touch on certain aspects of \texttt{pyPplusS}.
    
    Initial models for oblate and ringed exoplanets were based on 2D numerical integration. Albeit easy to implement, this approach is far too slow to be practical for large-scale searches. The first time, to our knowledge, that this approach was used in this context is in \citet{BarnesFortney03} and \citet{BarnesFortney04}. This approach is also used in other papers, \citep[e.g.][]{Tusnski11, Bourne18, Akinsanmi18}. Over the years, there were other techniques that are relevant, but are not directly related to modelling and simulation of light curves of exoplanets with rings:
    \begin{enumerate}	
    	\item \citet{Samsing15} discussed noise attenuation and in photometric data to allow the recovery of periodic features such as exoplanetary rings.
    	
    	\item \citet{Zuluaga15} developed a couple of potential criteria for candidate selection in searches for rings. Their analysis is aimed at candidate selection and initial vetting and not for positive detection and modelling of rings.
    	
        \item \citet[Section 6]{Visser15} discussed the phase curves of different types of planets as they spin about their axes, including ringed exoplanets.
        
    	\item \citet{Luger17} developed a model describing planet-planet occultations that uses similar terminology and ideas to the P+S model. In their analysis, they focused on complex limb darkening patterns (planetary illumination by the central star, in particular), however, they do not change the shape of the occulting body from a spherical planet, while we allow for oblate and ringed occulting bodies.
    
    \end{enumerate}
    
    We compared the results of \texttt{pyPplusS} and \texttt{exorings} in Figure \ref{fig:kenworthy_comp}. During the creation of this figure, we saw that \texttt{exorings} is significantly slower than \texttt{pyPplusS} and is prone to numerical errors dependant on the number of pixels used to calculate the star's illumination. Moreover, \texttt{exorings} only allows linear limb darkening, while our code can model non-linear limb darkening as well (See Section \ref{NonUniform}). Other comparable past works are of \citet{Aizawa17,Lecavelier17,Heising15} - unfortunately we could not find the source code of any of them. In \citet[Appendix A]{Aizawa17}, a slightly different algorithm for the ringed planet case is described. That algorithm, however, lacks the connection between uniform and non-uniform source calculations. Additionally, it requires multiple numerical integrals and thus it is more computationally demanding; our algorithm only requires a 1-D integral (see Appendix \ref{NumericalIntegration}). \citet[Section 5]{Lecavelier17} used a model for exoplanetary rings based on area calculations, but does not consider an inner radius for the ring and does not explain the model's handling of limb darkening. \citet{Heising15} used sectors to calculate a ring's contribution to the hidden area, but used an estimate for the star's local intensity which introduced modelling errors, hindering detection. They also checked just eight predetermined ring configurations rather than finding the best-fitting one.
    	
    To compare the speed of different algorithms, regardless of the machine on which they are run, we use \texttt{pyTransit} \citep{Parviainen15} -- an implementation of the MA02 model -- as a benchmark. Naturally, due to the much simpler geometry, it is significantly faster to run than ringed planets models. The speed of \texttt{pyPplusS} is roughly 2300 times slower than \texttt{pyTransit} on the same machine, while the performance quoted in \citet{Aizawa17}, which we believe to be the closest comparison to \texttt{pyPplusS}, is 3000 times slower than \texttt{pyTransit}. \texttt{pyPplusS} is therefore somewhat faster than the algorithm by \citet{Aizawa17}. However, note that the speed of the algorithm presented here depends on the system parameters and thus the speed may vary between different systems (See Appendix \ref{NumericalIntegration}). Importantly, the number of objects relevant for \texttt{pyPplusS} fitting, i.e. very high-SNR KOIs that are not eclipsing binaries, is limited to a few hundred. This is only a small fraction of the order of $10^{-2}$ of the systems examined in detail for spherical exoplanets (there were $\ge3.4 \cdot 10^4$ threshold crossing events in the final Kepler database). Since the computation time is about $10^{3}$ of the usual spherical planet fits, the total computational load is about one order of magnitude higher - which is very reasonable given that computers are now more capable than the ones available when \textit{Kepler} was launched in 2009.

    We note that a significant limitation of the computational efficiency of the algorithm in the case of ringed planets is the time needed for the high-precision solution of quartic equations, in order to calculate the positions of star--disc intersection points. That calculation takes more than 50 per cent of the current code's run time.

\section{Conclusions}
\label{Conclusions}

We developed the general, fast and precise Polygon+Segments (P+S) model, and described it specifically for the cases of light curves of oblate planets and planets with a ring. Such models are required in order to attempt detecting ringed exoplanets in existing \textit{Kepler} data or from future space-based transit surveys. The model also allows constraining the oblateness of planets using photometric data only, which is already available for many planets. The deviation caused by these effects from the standard MA02 model are small but possibly detectable with currently available data. The Polygon+Segments (P+S) model's core algorithm can be used to model light curves of more complicated configurations than before, including multiple planets, oblate planets, moons, rings, combinations thereof, etc., properly and efficiently taking into account overlapping areas and limb darkening. We make the P+S algorithm, implemented as a \texttt{pyPplusS} python package,  publicly available at CDS and at \href{https://github.com/EdanRein/pyPplusS}{GitHub}.

Some of the variables describing a ringed exoplanet are correlated in a simple way. For example, an increase of the planetary radius can be well compensated for by an increase of the ring's inner radius. We checked for correlations between all variables and found no unexpected correlation that could not be similarly understood.

\section*{Acknowledgements}


We would like to thank Prof. Oded Aharonson for many useful comments and suggestions. This project was supported by the Helen Kimmel Center for Planetary Science, the Minerva Center for Life Under Extreme Planetary Conditions and by the I-CORE Program of the PBC and ISF (Center No. 1829/12). AO acknowledges the support of the Koshland Foundation and McDonald-Leapman grant. This work was conducted as part of the Alpha Program at the Davidson Institute of Science Education, which is funded and operated nationally by Maimonides Fund's Center for the Advancement of the Gifted and Talented and the Education Ministry's Department for Gifted and Talented Students.

\bibliographystyle{plainnat}
\bibliography{MyBib}

\appendix

\section{Similar root of quartic equations}
\label{SimilarRoots}
    When using the quartic solver of \citet{Strobach10}, we found that the floating point errors introduced during the analytic start up of the algorithm sometimes caused divergence. This occurs when all of the solutions have a similar modulus, thus their order when sorted by their modulus is affected by numerical errors. When the solutions are two complex conjugate pairs with a similar modulus, $z_1,\Bar{z_1},z_2,\Bar{z_2}$, they might get sorted like this: $|z_1|<|z_2|<|\Bar{z_1}|<|\Bar{z_2}|$. Therefore, despite the existence of complex-conjugate pairs, the two value pairs from which the initialisation values for the chains are picked are not complex conjugates: $(z_2,\Bar{z_1}),(\Bar{z_1},\Bar{z_2})$. This causes the algorithm to not converge properly. Therefore when the maximum modulus difference between the analytic solutions is small ($<10^{-12}$), we choose to set them as complex conjugate pairs. Note that this fix relies (in part) on the assumption, which is true in this case, that the coefficients of the polynomial are real and thus if $z$ is a root, then its complex conjugate is also a root of the polynomial.

\section{Tolerance parameter}
\label{ToleranceParameter}
    The finite resolution of the representation of numbers in computers necessitates us to add a tolerance parameter $tol$ to the code (we use $tol=10^{-8}$) due to the following effects:
    \begin{legal}
        \item Points close to the edge: the algorithm requires us to determine whether points are inside curves or not. Since the exact positions of intersection points suffer from numerical errors, we use a tolerance parameter, effectively increasing (slightly) the curves by $tol$, in order to 'catch' intersection points that are supposed to be inside the curve but 'fell' out due to numerical errors in the calculation of their coordinates. For example, if a circle is the set of all points closer than $r$ to some point, then we implement it as closer than $r+tol$ for the above purpose. If an intersection point is falsely classified as inside a curve, the area resulting from the calculations is not affected.
        
        \item Triple intersections and curve classification: in step \ref{step:points_test} we determine which points are on the border of the DHA. Sometimes an intersection point is very close to another curve that is not one of the curves that intersect at that point. In these cases, when the P+S algorithm needs to select the curve on which to continue from that point, it has to consider all the relevant curves and not only the two curves whose intersection produced/defined the point during the initial computation stage. Therefore, if a curve is very close to a point, within $tol$, then we consider the point to be on that curve as well, i.e. the P+S algorithm will consider advancing to the next point along that curve as well (if possible). An example: if curves $A$ and $B$ intersect at $P_1$ and curves $B$ and $C$ intersect at $P_2$ it may be that $P_1$ and $P_2$ are so close they may as well be considered the same point. In such cases, when selecting a curve to continue along from point $P_1$ onward, the algorithm will also consider curve $C$ -- even though $P_1$ is formally only on curves $A$ and $B$.
        
        This is a generalisation to finite resolution of the exact triple intersection case - when a single point is exactly at the intersection of three curves simultaneously. 
        
        \item Very thin rings, etc.: some cases at the edges of the physical parameter space, e.g. if the inner and outer bounds of the ring will be nearly identical (the distance between them is close to or less than machine precision or $tol$)
        , may cause the algorithm to crash, because of errors originating in the calculation of intersection points. We note that such configurations (ring thickness of order $O(tol)$) are undetectable by any current, or foreseen, telescopes.
    \end{legal}

\section{Numerical Integration}
\label{NumericalIntegration}

    The numerical integration can be done using Gaussian Quadrature: integration using the interpolation polynomial that arises from points at the roots of the appropriate orthogonal polynomial. In this case, we use Gauss-Legendre quadrature, i.e. the orthogonal polynomials are Legendre polynomials \citep[see e.g.][]{GaussQuadRef}.
	In addition, following \citet{Aizawa17}, we split the integration interval at points where the underlying analytic function $F^e$ changes or may change. This occurs when the expression for the points on the border changes with the change of the integration parameter ($r$ or $t$ as defined above). In conclusion, there are 14 potential suspect points \citep[see also][]{Aizawa17}:
	\begin{enumerate}
	    \item Ring-Planet intersection (8 points or less)
	    \item Extrema of distance from the planet's edge to the star's centre (2 points or less)
	    \item Extrema of distance from the ring's inner edge to the star's centre (2 points or less)
	    \item Extrema of distance from the ring's outer edge to the star's centre (2 points or less)
	\end{enumerate}
	\texttt{pyPplusS} determines the location of these $\le 14$ points and sorts them according to their distance from the star's centre to create separate integration sub-intervals between them. In each sub-interval, the algorithm performs a fixed-order Gaussian Quadrature integration. Comparison with the aforementioned grid model showed that Gaussian Quadrature of order 10 yields errors less than $10^{-6}$.
	
	The performance of the numerical integration can be improved by noting that interior and exterior to the planet $A_\textrm{h}$ is constant and thus $F^e$ has a closed, analytic form. Therefore, we have a closed form for the function and the integral can be performed analytically in those sub-intervals. 
	
	 \textbf{Setting Gaussian Quadrature order n:} When the planet's radius may be bounded and the limb darkening parameters are known or approximated \textit{a priori}, which is actually the usual case, one may bound the potential modelling error. It is not strictly correct, but in general, the larger the interval of the integration the harder it is for the Gaussian Quadrature technique to produce a good polynomial approximations of the integrand. Therefore, the 'hardest' case for the model is a spherical planet with the largest radius allowed by the data, since in this case the integration interval will be long. On the other hand, if there are multiple curves (planet and ring) the algorithm will make more divisions to the integration interval, and thus the error will be reduced. Following this reasoning, one can generate the 'true' model curve by running the \texttt{pyPplusS} model once with a high order, e.g. $n=$20 or 30, which is relatively slow, and then bound the order $n$ that is necessary to produce sufficiently small modelling errors, for example, to fit a given data set. To do that, we test models generated using lower-order $n$ values relative to the 'true' curve above - until their differences are significantly smaller than the given data set. This will allow improving performance without exceeding any pre-set modelling error. This few-seconds process takes a negligible amount of time since it is performed only once per dataset. The code package \texttt{pyPplusS} includes a script implementing the procedure described above.
	
	To illustrate the effect of the numerical integration on the computational load, we assume a planet with $r_\textrm{p}=0.1,r_\textrm{in}=0.12,r_\textrm{out}=0.2,b=0.05$. For such a planet, the integration interval is split into $\sim6.7$ sub-intervals on average (depending on the planet's position). Therefore, calculation of the limb darkened light curve at $l$ equally spaced points during the transit the algorithm will call the uniform source function $\sim6.7\cdot n\cdot l$ times to produce one model light curve.

\end{document}